\documentclass[runningheads]{llncs}
\titlerunning{Behavioral Fingerprint-based Detection of Suspicious Account Misuse}
\usepackage[T1]{fontenc}

\usepackage{graphicx}

\usepackage{url}
\usepackage{amsmath}
\usepackage{amssymb}
\usepackage{booktabs}
\usepackage{multirow}
\usepackage{subcaption}

\begin{document}

\title{You Are Not My Teammate: \\ 
Behavioral Fingerprint-based Detection of Suspicious Account Misuse}

\author{Dong Hwan Lee \and Huy Kang Kim}
\institute{School of Cybersecurity, Korea University, Republic of Korea\\
\email{\{20-10076, cenda\}@korea.ac.kr}}

\maketitle           
\begin{abstract}
Online games have been continuously affected by cyber threats such as game bots and gold farming. Game bots, which are automated programs that play on behalf of human users, significantly accelerate character progression and reduce the engagement of legitimate players, potentially leading to user churn. In addition, gold farming enables the monetization of in-game currency into real-world money, resulting in unfair profits. For these reasons, prior studies have primarily focused on detecting game bots and gold farming. However, in competitive Multiplayer Online Battle Arena (MOBA) games such as League of Legends, match outcomes and rankings are the primary objectives, where individual performance is more critical than in-game economic factors. Accordingly, account misuse such as account sharing and boosting has emerged as a major threat to fair competition. 
In this study, we propose a behavioral fingerprint-based detection method. Our approach analyzes and quantifies changes between a player's historical and recent in-game behaviors. Consequently, it enables the robust identification of suspicious account sharing and boosting, even in label-scarce environments.
Experimental results show that behavioral fingerprints within the same account are distinguishable from those across different accounts, supporting rapid detection of suspicious account misuse even with limited labeled data.

\end{abstract}

\section{Introduction}
\begin{figure}
    \centering
    \includegraphics[width=\textwidth]{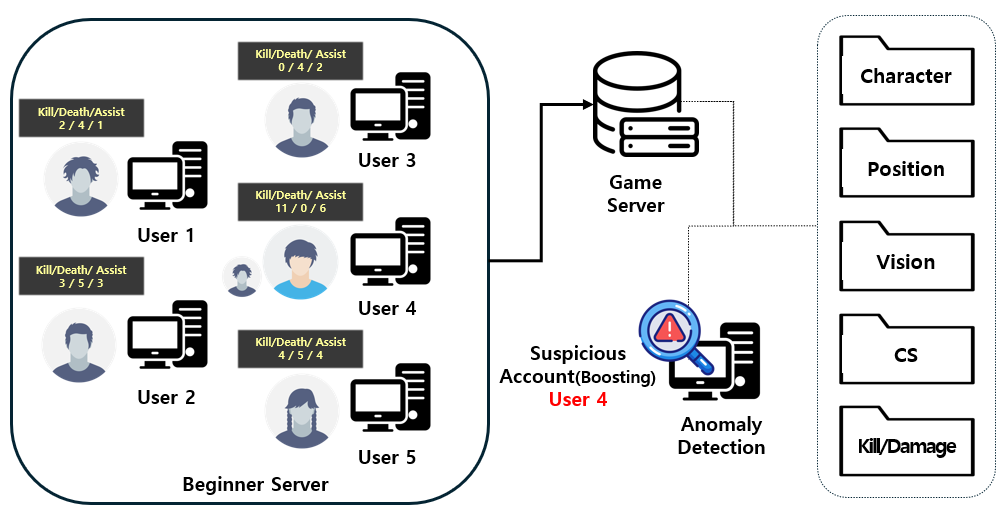}
    \caption{Overview of the proposed approach for detecting 
suspicious account misuse based on behavioral fingerprint analysis.}
    \label{fig1}
\end{figure}

Online games have evolved beyond entertainment into major social and economic platforms, giving rise to diverse cheating behaviors including game bots and gold farming \cite{a1}. However, in competitive Multiplayer Online Battle Arena (MOBA) games such as League of Legends, individual skill and rank are more critical than in-game economic assets, making account misuse such as account sharing and boosting major concerns \cite{a2}. In particular, when a player with abnormally high skill intervenes in a matchmaking environment intended for similarly skilled players, the balance of the game is disrupted. 
This inflicts repeated, unfair losses on legitimate players, directly degrading user engagement and accelerating churn \cite{a3}.
Thus, detecting and mitigating account misuse is critical in MOBA games from the perspective of game's security and fairness. To address these issues, Riot Games explicitly prohibits account misuse in its terms of service and states that violations may result in penalties such as service restrictions \cite{a4}. This indicates that the threats posed by boosting are recognized not as a regional concern, but as a universal challenge to competitive fairness in online gaming. Conventional login-based security mechanisms primarily focus on authentication of users and detection of abnormal login activities.  
However, after authentication, it is difficult to directly verify whether the actual player is the legitimate account owner. Furthermore, since boosting is performed by human players rather than game bots, it is difficult to detect using conventional approaches such as social profiling. To address these challenges, we propose an approach that detects suspicious account misuse based on players' gameplay fingerprints, focusing on the fact that a player's gameplay style and strategy shift when a different player controls the account. 
Fig. \ref{fig1} illustrates the process of detecting suspicious activities based on changes in behavioral fingerprints caused by a player change. We apply a normalization technique that accounts for structural differences across positions, thereby reducing bias and effectively capturing detailed behavioral variations within the same position. Furthermore, by comparing past and recent behavioral fingerprints of a player, we can effectively identify suspicious accounts even in a label-scarce environment. The main contributions of this paper are as follows:
\begin{itemize}
\item We propose a behavioral fingerprint-based method that enables rapid triage of suspicious account misuse using only server-side match logs.
\item We apply role-based normalization to reduce structural bias across positions and enable fair behavioral comparison.
\item We show that intra-account behavioral fingerprints are distinguishable from inter-account fingerprints, supporting detection in label-scarce environments.
\end{itemize}

\section{Related Work}
\subsection{Online Game Security and Cheating}
With the growth of online games, various forms of cheating have continuously increased, leading to the development of diverse detection methods. Early studies systematically categorized cheating in online games and defined fairness as a core element of game security \cite{a5}, followed by research that further classified cheating in MMORPG environments into game bots, account theft, and gold farming, along with corresponding detection methods \cite{a6}. Subsequent studies detected game bots using multimodal behavioral features \cite{a7} and self-similarity \cite{a8}, and identified gold farming networks via trade graph analysis \cite{a9}. These studies primarily focus on detecting cheating behaviors driven by automation or economic profit. 
However, beyond cheating driven by automation or economic profits, account misuse such as account sharing and boosting has also been identified as a critical threat to game fairness, and prior studies have analyzed the concept of boosting and its negative impacts, including unfair matchmaking and the degradation of ranking system integrity \cite{a2,a10}. Nevertheless, studies that directly detect boosting using gameplay data remain limited. In MOBA games, particularly in League of Legends, prior research has explored matchmaking fairness and rank system design \cite{a10,a11}, and player performance evaluation and match outcome prediction \cite{a12,a13}. However, despite this breadth of League of Legends research, no prior study has directly addressed the detection of account misuse such as account sharing and boosting using server-side match logs alone. This gap motivates the present work. 

\subsection{Behavioral and Biometric-based Approaches}
Research aimed at identifying individuals based on players’ behavioral patterns has also been continuously conducted. Early studies showed that player-specific characteristics could be captured through idle-time distributions during gameplay \cite{a14}. Subsequently, an approach was proposed that modeled the process of identity theft step by step and detected anomalous behavior by combining behavioral patterns with economic activities \cite{a15} or by analyzing server-side action sequences \cite{a23}. 
In FPS environments, behavioral fingerprints such as aim accuracy and movement patterns have been used to detect cheating \cite{a16}, demonstrating that server-side behavioral analysis is feasible.
User authentication using mouse and keystroke dynamics has also been explored, including account owner identification in League of Legends \cite{a17}. However, such approaches are constrained in real-world service deployment because they require client-side data collection. Accordingly, we complement prior work by detecting suspicious account misuse using only server-side match logs.

\section{Background}
\subsection{Account Misuse}
Account misuse refers to the use of a game account by an individual other than the legitimate owner, in violation of the platform's Terms of Service \cite{a4}. Two primary forms exist in competitive MOBA environments: account sharing, 
where another individual uses someone else's account, and boosting, where a highly skilled player plays on behalf of another user's account to artificially increase its rank. Both behaviors are distinct from account theft and cause a 
noticeable shift in the account's behavioral fingerprint regardless of motivation. Since the goal of this work is to detect accounts that show such a shift, our method does not need to distinguish between the two cases. 

\subsection{Position-based Game Structure and Early Game}
In League of Legends, each player is assigned one of five positions — Top, Jungle, Mid, Bottom, and Support — per match, each with distinct roles and behavioral patterns. For example, Bottom and Jungle tend to focus on Creep Score (CS), whereas Support emphasizes vision control and team combat \cite{a12}. Due to these structural differences, position-based bias must be accounted for when analyzing behavioral fingerprints. 

Similarly, the gameplay evolves over time from an early phase focused on individual play to a late phase dominated by team combat. The period up to approximately 10 minutes after the start of a match is defined as the early game, and it can have a significant impact on match outcomes \cite{a18}. During this phase, players primarily engage in individual play, allowing their personal play styles to be directly reflected. Due to these characteristics, early game data are suitable for analyzing players' distinctive behavioral features.

\subsection{Behavioral Fingerprint}
A behavioral fingerprint refers to a quantitative representation of a player's distinctive play style and strategy, based on the concept that individuals can be distinguished through their in-game behavior. In MOBA environments, this encompasses aspects such as champion selection preferences, farming behavior, and combat participation patterns, which inherently change when the actual player changes.

\section{Method}
\subsection{Data Collection}
We collected match data via the Riot Games API by randomly selecting seed accounts from ranked players on the Korean server and iteratively expanding the account set through match participant information. Since our proposed approach analyzes behavioral patterns by comparing windows of consecutive matches within each account, sufficient per-account match records are essential to construct both the baseline and recent windows reliably. Because the number of available matches varied considerably across accounts, we selectively collected additional matches for accounts with insufficient records in later stages. We excluded accounts with fewer than 35 matches, yielding 100 accounts for analysis. 

\subsection{Feature Design}
To design behavioral features, we extracted server-side match and timeline metrics from the Riot Games API and constructed derived features to capture behavioral changes within an account. Based on prior MOBA and League of Legends studies~\cite{a10,a12,a13,a18}, we extracted 12 derived features from server-side match logs and organized them into five categories: champion selection pattern, position distribution, farming behavior, vision score, and combat participation behavior. We computed farming behavior from the first 10 minutes of each match to capture individual early-game behavior. For features that vary significantly across positions, we applied role-based Z-score normalization.

\subsubsection{Champion Selection Pattern}
Champion selection pattern captures a player's champion preference using Champion Pool Size and Champion Usage Entropy. Champion Pool Size measures the number of distinct champions used, while Champion 
Usage Entropy captures how evenly champion selections are distributed:
\begin{equation}
\textit{ChampionPoolSize} = |\{\textit{Champion}_k : k = 1, \ldots, N\}|, \quad
H = -\sum_{i=1}^{C} p_i \log p_i
\end{equation}
where $\textit{Champion}_k$ denotes the champion played in the $k$-th match, and $N$ is the total number of matches in the window. $C$ denotes the number of distinct champions, $p_i$ denotes the proportion of matches in which the $i$-th distinct champion was played, and $H$ denotes the champion usage entropy. 

\subsubsection{Position Distribution}
Unlike champion selection, which spans a large and variable set of options, position is drawn from a small fixed set of five roles, allowing a direct proportion-based formulation. Position Distribution represents a player's position preference as a 
five-dimensional vector, where each dimension corresponds to the proportion of matches played in one of the five positions:
\begin{equation}
P_r
=
\frac{1}{N}
\sum_{i=1}^{N}
\mathbb{I}(pos_i = r)
\label{eq:position_distribution}
\end{equation}
where $N$ is the total number of matches, $\mathit{pos}_i$ denotes the position selected by the player in the $i$-th match, $r \in \{\text{TOP, JUNGLE, MIDDLE, BOTTOM, SUPPORT}\}$, and $\mathbb{I}$ is the indicator function.

\subsubsection{Farming Behavior}
We selected CS-based features to represent a player's farming ability. To more accurately capture individual behavior, we used CS per minute at 10 minutes. Since CS distributions differ substantially across positions, we applied role-based Z-score normalization: 
\begin{equation}
\mathit{CS}_{10} = \frac{\mathit{CS\ at\ 10\ min}}{10}, \quad
\mathit{CS}_{\mathit{role\_z}} = 
\frac{\mathit{CS}_{10} - \mu_{\mathit{role}}}{\sigma_{\mathit{role}}}
\end{equation}
To capture both performance and consistency across matches, we used $\mathit{CS}_{\mathit{role\_z\_mean}}$ and $\mathit{CS}_{\mathit{role\_z\_std}}$ as derived features.

\subsubsection{Vision Score}
Vision Score captures vision control behavior, including ward placement and removal. To remove the effect of match duration and position-based differences, we first divided the raw Vision Score $V$ by the match duration $t$ (in minutes) and then normalized it within each position:
\begin{equation}
\mathit{Vision}_{\mathit{role\_z}} = 
\frac{(V / t) - \mu_{\mathit{role}}}{\sigma_{\mathit{role}}}
\end{equation}

\subsubsection{Combat Participation Behavior}
Combat Participation Behavior was represented using Kill Participation (KP) and Damage Share (DS). KP measures a player's involvement in team kills, while DS measures the proportion of total team damage dealt by the player. We normalized both features within each position to reduce position-based bias:
\begin{equation}
\mathit{KP} = \frac{\mathit{Kills} + \mathit{Assists}}
{\mathit{Team\ Kills}}, \quad
\mathit{DS} = \frac{\mathit{Damage\ Dealt}}
{\mathit{Team\ Total\ Damage}}
\end{equation}
\begin{equation}
\mathit{KP}_{\mathit{role\_z}} = 
\frac{\mathit{KP} - \mu_{\mathit{role}}}{\sigma_{\mathit{role}}}, 
\quad
\mathit{DS}_{\mathit{role\_z}} = 
\frac{\mathit{DS} - \mu_{\mathit{role}}}{\sigma_{\mathit{role}}}
\end{equation}

\subsection{Behavioral Fingerprint Modeling}
To quantitatively model player behavior, we constructed behavioral fingerprints for each account based on match data ordered chronologically. We generated windows consisting of a fixed number 
of matches, each represented as a single 12-dimensional vector. For example, we summarized kill participation and damage share as their mean across matches in the window. We defined a baseline window to represent a stable gameplay style and a recent window to capture changes in behavioral fingerprints. 

The baseline window used a relatively larger window size to reflect long-term behavioral patterns, whereas the recent window used a smaller window size to sensitively capture short-term shifts in gameplay style. We set candidate window sizes to 5, 10, 20, 25, and 30 matches, and the selection criteria are described in Section 5.3.

\subsection{Anomaly Detection}
\begin{figure}[t]
    \centering
    \includegraphics[width=0.8\linewidth]{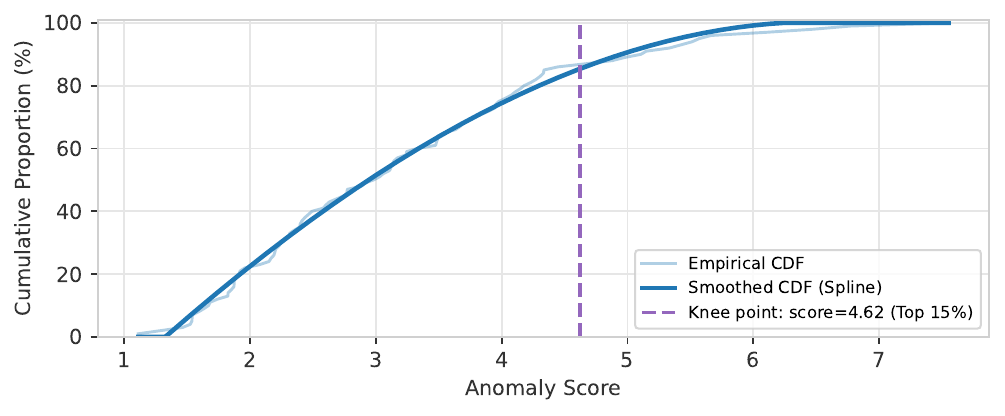}
    \caption{Empirical CDF of anomaly scores. The Kneedle threshold is detected at score 4.62.}
    \label{fig:cdf_threshold}
\end{figure}

To identify suspicious account misuse, we defined the change in behavioral fingerprints between the baseline and recent windows as an anomaly score \cite{a19}, computed as the Euclidean distance between their feature vectors:
\begin{equation}
d = \left\| v_{\mathit{baseline}} - v_{\mathit{recent}} 
\right\|_2
\end{equation}
where $v_{baseline}$ and $v_{recent}$ denote the feature vectors of each window. Prior to computation, we applied StandardScaler to ensure that no single feature disproportionately influences the distance. To determine the threshold in a data-driven manner, we applied the Kneedle algorithm~\cite{a20} to the empirical CDF of anomaly scores, which identifies the knee point marking the boundary 
between the concentrated normal region and the sparse upper tail directly from the data without requiring an arbitrarily chosen 
percentile. We fitted a cubic smoothing spline to the CDF with the initial value set to zero to satisfy the boundary condition. We set the sensitivity parameter $S$ to 1 to avoid premature knee detection, confirming consistency for $S = 2$ and $S = 3$. The resulting threshold at score 4.62 corresponded to the top 15\%, as shown in Fig.~\ref{fig:cdf_threshold}. We therefore classified accounts with anomaly scores above this threshold as suspicious~\cite{a21}.

\section{Experimental Design and Analysis}
\subsection{Experimental Setup}
We collected data from January to March 2026 using the Riot Games API, targeting ranked solo queue matches on the League of Legends Korean server. We selected the Korean server because it provides an active and highly competitive ranked environment, and a prior study has also reported account sharing sanctions in the Korean competitive gaming context \cite{a2}. We excluded accounts with fewer than 35 matches to ensure sufficient data for simultaneously constructing candidate baseline windows of up to 30 matches and a recent window of 5 matches, yielding 100 accounts for analysis. To examine whether the proposed method responds to behavioral changes, we conducted a controlled behavioral-change validation using 10 accounts. In each case, the original account owner intentionally altered their gameplay style and strategy across a minimum of 5 consecutive ranked matches while playing the account personally. For example, we instructed participants to adopt a more passive combat style than usual while focusing more aggressively on farming. This validation does not constitute actual account sharing or boosting, but serves to examine whether the proposed behavioral fingerprint can detect controlled behavioral changes within an account.

\subsection{Validity Verification}
We conducted validity verification to confirm the core assumption underlying the proposed detection approach — namely, that distances between behavioral fingerprints within the same account are significantly smaller than those between different accounts. For each account, we set the most recent 35 matches as the reference interval and constructed nested windows of sizes 5, 10, 20, 25, 30, and 35 matches, all anchored at the most recent match within the reference interval. We represented each window as a 12-dimensional behavioral fingerprint vector comprising features such as champion selection diversity and position distribution. We compared intra-account distances and inter-account distances, following the same standardization procedure described in Section 4.4. Table \ref{tab:validity} summarizes the representative intra- and inter-account distances. Even the largest intra-account distance (3.293 for 5 vs. 35) remained substantially below the smallest inter-account distance (4.489 for 30 vs. 30), confirming a clear separation between the two distributions. These findings demonstrate that behavioral fingerprints stably capture account-level patterns and are applicable to detecting suspicious account misuse.

\begin{table}[t]
\centering
\caption{Intra-account and Inter-account Distance Comparison}
\label{tab:validity}
\renewcommand{\arraystretch}{0.91}
\setlength{\tabcolsep}{18pt}
\begin{tabular}{|c|c|c|c|}
\hline
\textbf{Distance Type} & \textbf{Window Pair} 
& \textbf{Mean} & \textbf{Median} \\
\hline
\multirow{5}{*}{\shortstack{Intra-account\\distance\\(maximum per\\anchor window)}}
& 5 vs 35  & 3.293 & 3.025 \\
& 10 vs 35 & 2.285 & 2.197 \\
& 20 vs 35 & 1.282 & 1.243 \\
& 25 vs 35 & 0.875 & 0.796 \\
& 30 vs 35 & 0.509 & 0.444 \\
\hline
\multirow{6}{*}{\shortstack{Inter-account\\distance}}
& 5 vs 5   & 5.179 & 5.105 \\
& 10 vs 10 & 4.747 & 4.791 \\
& 20 vs 20 & 4.496 & 4.521 \\
& 25 vs 25 & 4.492 & 4.513 \\
& 30 vs 30 & 4.489 & 4.488 \\
& 35 vs 35 & 4.513 & 4.500 \\
\hline
\end{tabular}
\end{table}

\subsection{Baseline Window Size Selection}
\begin{figure}[t]
    \centering
    \begin{subfigure}[b]{0.48\textwidth}
        \centering
        \includegraphics[width=\textwidth]{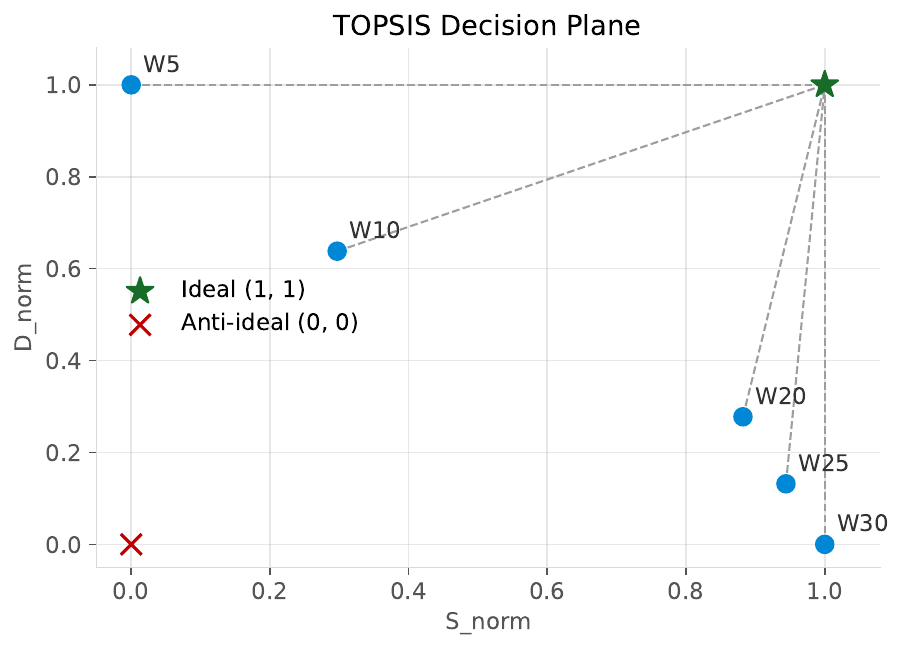}
        \caption{Decision Plane}
        \label{fig:topsis_plane}
    \end{subfigure}
    \hfill
    \begin{subfigure}[b]{0.48\textwidth}
        \centering
        \includegraphics[width=\textwidth]{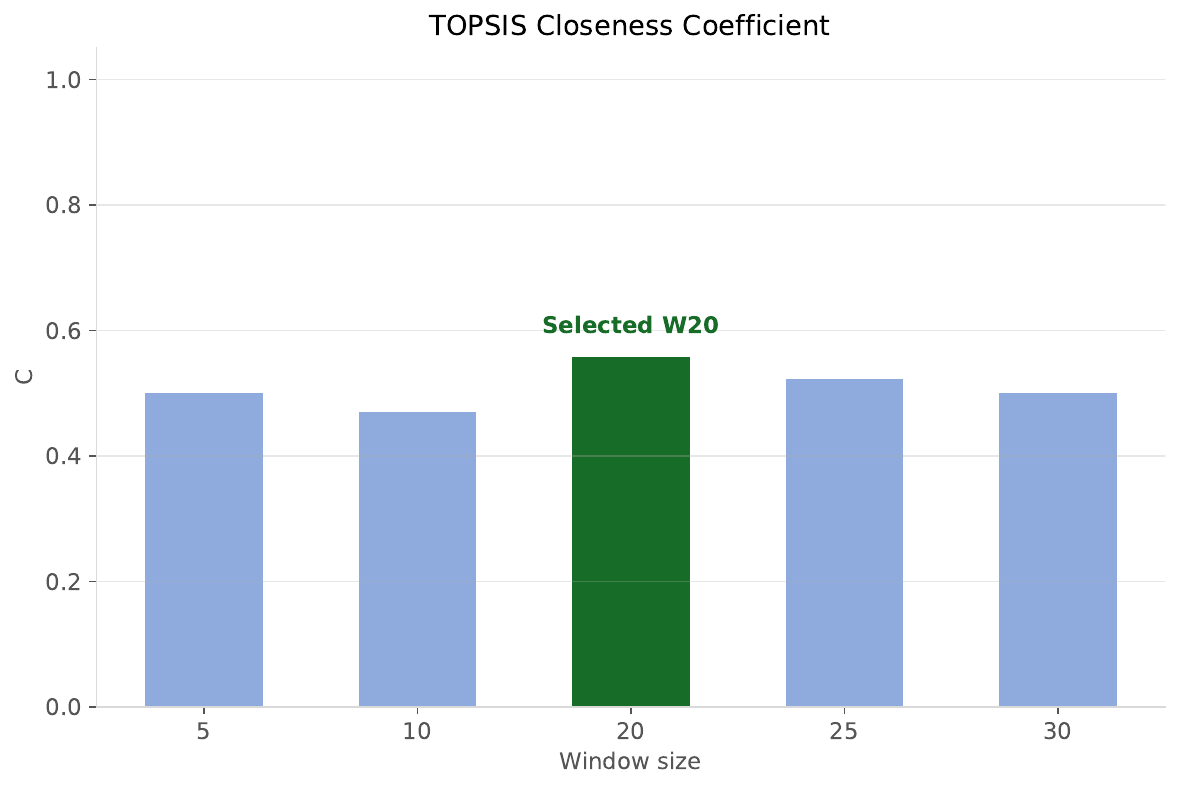}
        \caption{Closeness coefficient $C_w$ per window}
        \label{fig:topsis_coeff}
    \end{subfigure}
    \caption{TOPSIS-based baseline window size selection. (a) Each candidate window is plotted in the normalized criterion space. (b) The 20-match window achieves the highest closeness coefficient and is selected as the baseline window.}
    \label{fig:topsis}
\end{figure}

To determine the optimal window sizes, we measured two 
criteria for each candidate window. Stability ($S_{\text{norm}}$) measures how consistently a behavioral fingerprint is maintained when transitioning to an adjacent larger window, measured as the Euclidean distance between adjacent window pairs, with inverted Min-Max normalization applied so that higher scores correspond to greater stability. Sensitivity ($D_{\text{norm}}$) measures how well a window captures the difference between short-term and long-term behavioral patterns, measured as the Euclidean distance between each window and the maximum-size window of 35 matches, normalized via standard Min-Max normalization. To objectively select the optimal baseline window size, we employed TOPSIS \cite{a22} with normalized stability and sensitivity as evaluation criteria. Letting $\mathbf{v}_{w} = (S_{\text{norm},w},\, D_{\text{norm},w})$ denote the criterion vector of window $w$, we defined the ideal solution $\mathbf{v}^{+} = (1, 1)$ and anti-ideal solution $\mathbf{v}^{-} = (0, 0)$. We computed the distance to each reference point and the closeness coefficient $C_w$ as:
\begin{equation}
d^{\pm}_{w} = \lVert \mathbf{v}_{w} - \mathbf{v}^{\pm} 
\rVert_{2}, \qquad
C_{w} = \frac{d^{-}_{w}}{d^{+}_{w} + d^{-}_{w}}
\end{equation}
As shown in Fig. \ref{fig:topsis}, the 20-match window achieved the highest closeness coefficient ($C = 0.558$), so we selected it as the baseline window. Since we considered fewer than 5 matches likely to compromise statistical stability, we established 5 matches as the minimum for the recent window.

\section{Results and Discussion}

\subsection{Suspicious Account Detection Results}
\begin{figure}[t]
    \centering
    \begin{subfigure}[b]{0.48\textwidth}
        \centering
        \includegraphics[width=\textwidth]{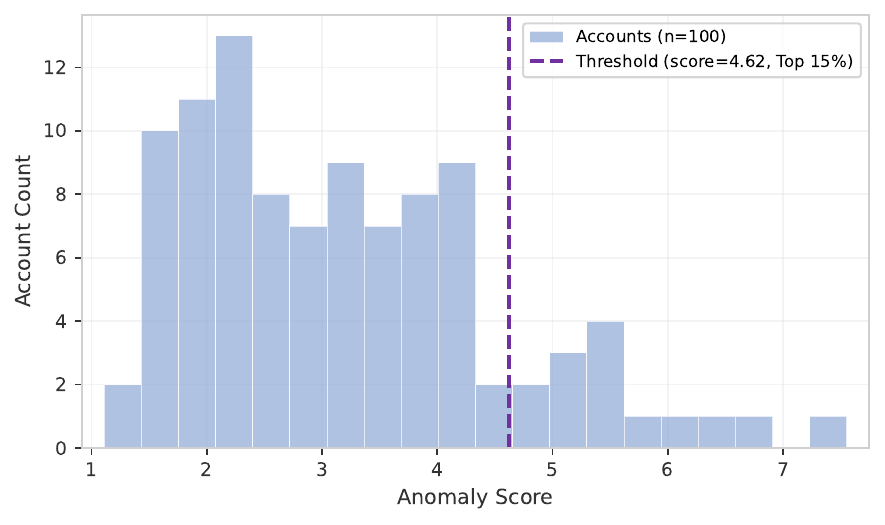}
        \caption{Anomaly scores (100 accounts)}
        \label{anomaly_dist1}
    \end{subfigure}
    \hfill
    \begin{subfigure}[b]{0.48\textwidth}
        \centering
        \includegraphics[width=\textwidth]{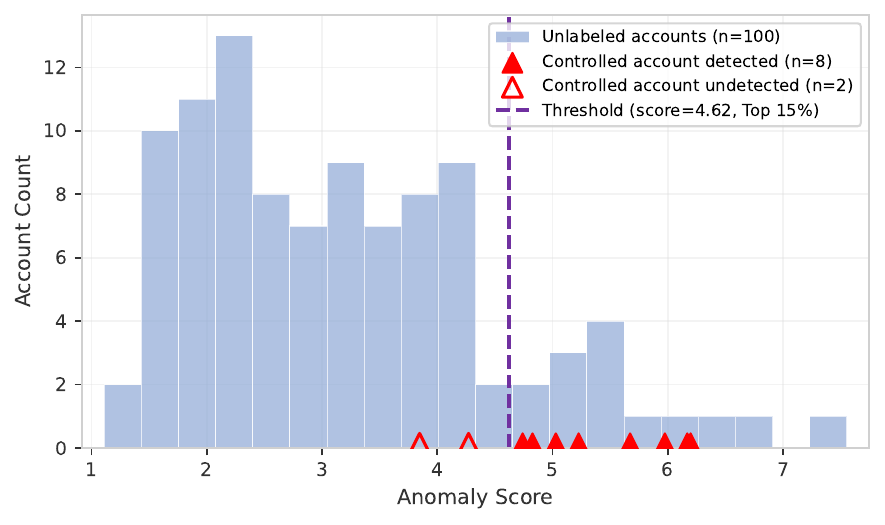}
        \caption{Anomaly scores (Controlled accounts)}
        \label{anomaly_dist2}
    \end{subfigure}
    \caption{Distribution of anomaly scores with threshold at score 4.62 (dashed line). (a) Most accounts concentrate in the lower region with few in the upper tail. (b) Among 10 controlled accounts (red triangles), 8 are detected and 2 are not.}
    \label{fig:anomaly_dist}
\end{figure}

We computed anomaly scores based on a 20-match baseline window and a 5-match recent window, with the baseline constructed from the 20 matches immediately preceding the recent window. We present the score distribution across 100 accounts in Fig. \ref{fig:anomaly_dist}(a), ranging from 1.11 to 7.55 with a mean of 3.19 and a median of 3.01. Applying the threshold of 4.62, we classified 15 accounts (the top 15\%) as suspicious. To validate detection capability, we introduced 10 controlled behavioral-change accounts into the pipeline. As shown in Fig. \ref{fig:anomaly_dist}(b), 8 of the 10 controlled accounts (80\%) exceeded the threshold. The two undetected accounts recorded scores of 4.27 and 3.85, which may reflect a relatively moderate degree of behavioral change compared to more pronounced real-world cases. Our feature-level analysis of the detected accounts revealed that $\mathit{CS}_{\mathit{role\_z\_std}}$ and position distribution consistently exhibited the largest variation, emerging as primary indicators of a gameplay style shift. 
These results indicate that our approach can effectively identify suspicious accounts in a label-scarce environment, demonstrating its potential as a rapid triage tool for further analysis.

\subsection{Discussion}
While our proposed approach demonstrated the ability to effectively identify suspicious account misuse, several limitations warrant consideration. First, we conducted experiments on 100 accounts, which may not be sufficient to validate generalizability, and future work should construct larger-scale datasets. We instantiated our framework using League of Legends features, and testing it on other games remains future work as well. Second, since we explicitly instructed participants to alter specific behaviors, the observed changes may partly reflect these instructions, and this controlled setting may not fully capture the degree of gameplay style shifts observed in actual misuse cases. Finally, we defined the baseline window as the 20 matches immediately preceding the recent window. If account misuse persists over an extended period, behavioral patterns from the misuse period may be incorporated into the baseline window, diluting the detectable change and limiting the method to short-term or sporadic misuse rather than long-term, persistent cases. 
Future research may consider dynamically configuring the baseline window to mitigate this issue. While false positives cannot be fully excluded, we designed the proposed approach to serve as a rapid triage tool, and we treat accounts exceeding the threshold as suspicious candidates for further analysis, not as confirmed misuse cases.

\section{Conclusion}
We proposed a behavioral fingerprint-based detection method for identifying suspicious account misuse such as account sharing and boosting in MOBA game environments. Unlike prior studies focused on game bot detection or player identification, we addressed the problem of detecting behavioral changes within the same account using a sliding-window framework and an unsupervised approach to ensure applicability in label-scarce environments. 
Our experimental results confirmed that intra-account distances were significantly smaller than inter-account distances, and the data-driven threshold derived from the Kneedle algorithm identified 15 suspicious accounts in the upper tail of the score distribution. Our validation using 10 controlled accounts further demonstrated an 80\% detection rate. These results suggest that our proposed approach can serve as a rapid triage tool, enabling early detection of suspicious accounts. Among the features, we identified $\mathit{CS}_{\mathit{role\_z\_std}}$ and position preference as primary indicators of play style shift. However, the detection results on unlabeled accounts remain at the level of suspicion, as verified labels for real-world misuse cases were unavailable. Future work will validate detection performance using larger datasets with verified labels and extend the framework to handle long-term misuse scenarios.

\subsubsection*{Acknowledgements.}
This work was supported by Korea Internet \& Security Agency (KISA) 
grant funded by the Korea government (PIPC) 
(No. 2780000036, Masters and Doctoral Program for Advanced Talent 
Development in Personal Data Protection and Responsible Use).

\bibliographystyle{splncs04}


\end{document}